\documentclass{article}
\usepackage{spconf,amsmath,graphicx}

\usepackage{color,xcolor}
\usepackage{booktabs}
\usepackage{xspace}
\usepackage{newtxmath}
\usepackage{rotating}
\usepackage{multirow}
\usepackage{balance}
\usepackage{url}
\newlength\savewidth
\usepackage[colorlinks,citecolor=citecolor, linkcolor=linkcolor, bookmarks=false]{hyperref}
\definecolor{citecolor}{HTML}{1F801F}
\definecolor{linkcolor}{HTML}{ED1C24}
\usepackage{collab}
\collabAuthor{shan}{red}{HS'Note}
\collabAuthor{yjx}{orange}{JX'Note}
\collabAuthor{wxc}{orange}{XC'Note}

\newcommand{\methodname}{TIM-Net\xspace}
\newcommand{\T}{^{\textrm T}} 

\newcommand{\vct}[1]{\boldsymbol{#1}} 
\newcommand{\mat}[1]{\boldsymbol{#1}} 
\newcommand{\etal}{\textit{et al}.\xspace}
\newcommand{\ie}{\textit{i}.\textit{e}.,\xspace}
\newcommand{\eg}{\textit{e}.\textit{g}.,\xspace}
\newcommand{\tabincell}[1]{\begin{tabular}[l]{@{}c@{}} #1\end{tabular}}
\newcommand{\addline}[1]{\multicolumn{1}{c|}{#1}}
\makeatletter
\DeclareRobustCommand{\cev}[1]{%
  {\mathpalette\do@cev{#1}}%
}
\newcommand{\do@cev}[2]{%
  \vbox{\offinterlineskip
    \sbox\z@{$\m@th#1 x$}%
    \ialign{##\cr
      \hidewidth\reflectbox{$\m@th#1\vec{}\mkern4mu$}\hidewidth\cr
      \noalign{\kern-\ht\z@}
      $\m@th#1#2$\cr
    }%
  }%
}

\title{Temporal Modeling Matters: A Novel Temporal Emotional Modeling Approach for Speech Emotion Recognition}
\name{Jiaxin Ye$^1$, Xin-cheng Wen$^2$, Yujie Wei$^1$, Yong Xu$^3$, Kunhong Liu$^{4,\dagger}$, Hongming Shan$^{1,5,\dagger}$\thanks{$\dagger$: Co-corresponding author.}}
\address{$^1$Institute of Science and Technology for Brain-Inspired Intelligence, Fudan University, Shanghai, China\\
 $^2$Department of Computer Science, Harbin Institute of Technology (Shenzhen), Shenzhen, China\\
 $^3$School of Computer Science and Mathematics, Fujian University of Technology, Fuzhou, China\\
$^4$School of Film, Xiamen University, Xiamen, China\\
$^5$Shanghai Center for Brain Science and Brain-Inspired Technology, Shanghai, China
}

\begin{document}
%
\maketitle
\begin{abstract}
Speech emotion recognition (SER) plays a vital role in improving the interactions between humans and machines by inferring human emotion and affective states from speech signals. 
Whereas recent works primarily focus on mining spatiotemporal information from hand-crafted features, we explore how to model the temporal patterns of speech emotions from dynamic temporal scales.
Towards that goal, we introduce a novel temporal emotional modeling approach for SER, termed \textbf{T}emporal-aware b\textbf{I}-direction \textbf{M}ulti-scale Network (\textbf{\methodname}), which learns multi-scale contextual affective representations from various time scales. 
Specifically, \methodname first employs temporal-aware blocks to learn temporal affective representation, then integrates complementary information from the past and the future to enrich contextual representations, and finally fuses multiple time scale features for better adaptation to the emotional variation.
Extensive experimental results on six benchmark SER datasets demonstrate the superior performance of \methodname, gaining 2.34\% and 2.61\% improvements of the average UAR and WAR over the second-best on each corpus. 
The source code is available at \url{https://github.com/Jiaxin-Ye/TIM-Net_SER}.

\end{abstract}


\section{INTRODUCTION}

\begin{figure*}[t]
	\centering
	\includegraphics[width=1.0\textwidth]{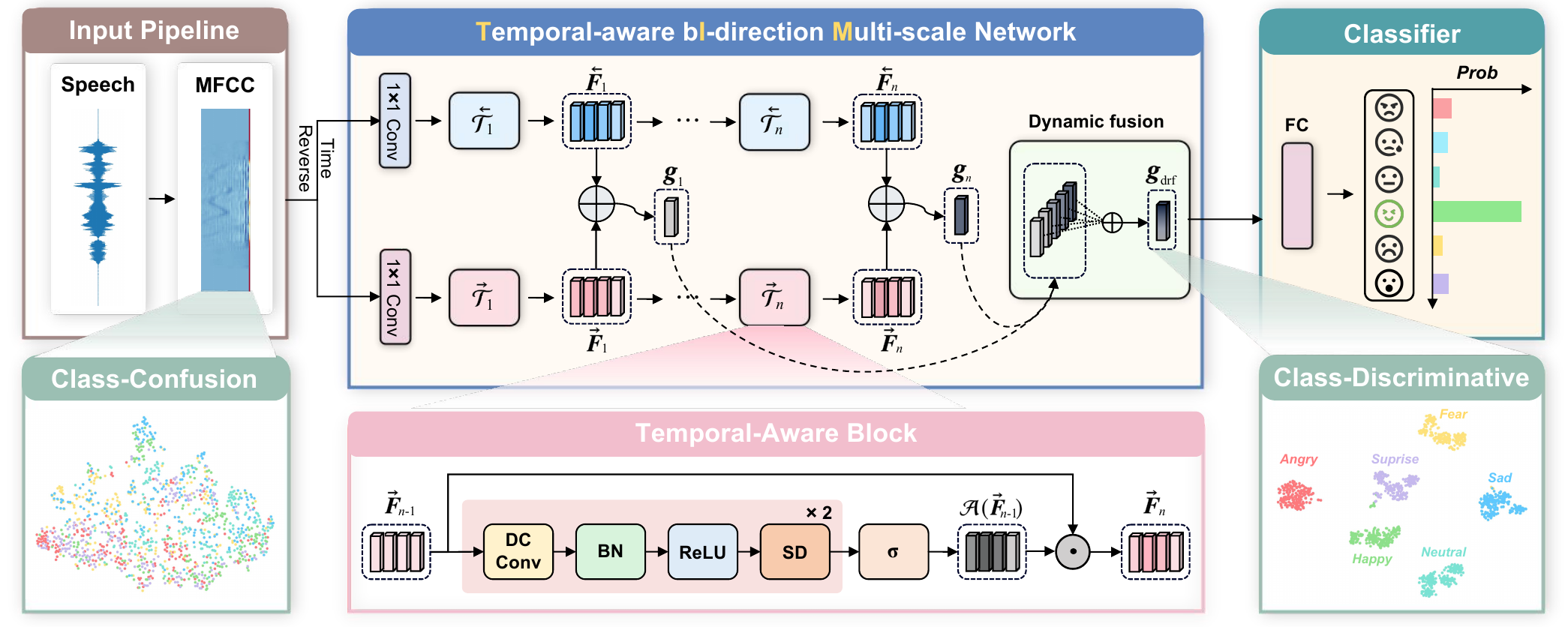}
	\caption{The framework of the \methodname for learning affective features, whose feature extraction part is composed of a bi-direction module and a dynamic fusion module. Note that the forward $\vec{\mathcal{T}}_{j}$ and backward $\cev{\mathcal{T}}_{j}$ are the same structure with different inputs.}
	\label{fig:architecture}
\end{figure*}

Speech emotion recognition (SER) is to automatically recognize human emotion and affective states from speech signals, enabling machines to communicate with humans emotionally~\cite{schuller2018speech}. It becomes increasingly important with the development of the human-computer interaction technique. 

The key challenge in SER is \emph{how to model emotional representations from speech signals}. 
Traditional methods~\cite{TSP_INCA,INCA_CNN} focus on the efficient extraction of hand-crafted features, which are fed into conventional machine learning methods, such as Support Vector Machine (SVM).
More recent methods based on deep learning techniques aim to learn the class-discriminative features in an end-to-end manner, which employ various architectures such as Convolutional Neural Network (CNN)~\cite{RM_CNN, DBLP:conf/icpr/WenLZJ20/Capsule}, Recurrent Neural Network (RNN)~\cite{GRU_ser1,Dual_LSTM}, or the combination of CNN and RNN~\cite{bilstm_ser3}. 

In particular, various temporal modeling approaches, such as Long Short-Term Memory (LSTM), Gate Recurrent Unit (GRU), and Temporal Convolution Network (TCN), are widely adopted in SER, aiming to capture dynamic temporal variations of speech signals. For example, Wang~\etal~\cite{Dual_LSTM} proposed a dual-level LSTM to harness temporal information from different time-frequency resolutions. 
Zhong~\etal~\cite{IEMOCAP_GRU} used CNN with Bi-GRU and focal loss for learning integrated spatiotemporal features. 
Rajamani~\etal~\cite{GRU_ser1} presented an attention-based ReLU within GRU to capture long-range interactions among the features. 
Zhao~\etal~\cite{bilstm_ser3} leveraged fully CNN and Bi-LSTM to learn the spatiotemporal features. 
However, these methods suffer from the following drawbacks: 1) they lack sufficient capacity to capture long-range dependencies for context modeling, where the capture of the context in speech is crucial for SER since human emotions are usually highly context-dependent; and 2) they do not explore the dynamic receptive field of the model, while learning dynamic instead of maximal ones can improve model generalization ability to unknown data or corpus.

To overcome these limitations in SER, we propose a \textbf{T}emporal-aware b\textbf{I}-direction \textbf{M}ulti-scale Network, termed \textbf{\methodname}, which is a novel temporal emotional modeling approach to learn multi-scale contextual affective representations from various time scales. The contributions are three-fold. 
\emph{First}, we propose a temporal-aware block based on the Dilated Causal Convolution (DC Conv) as a core unit in \methodname. The dilated convolution can enlarge and refine the receptive field of temporal patterns. The causal convolution combined with dilated convolution can help model relax the assumption of first-order Markov property compared with RNNs~\cite{Markov}. In this way, we can incorporate an $N$-order ($N$ denotes the number of all previous frames) connection into the network to aggregate information from different temporal locations. 
\emph{Second}, we devise a novel bi-direction architecture integrating complementary information from the past and the future for modeling long-range temporal dependencies. To the best of our knowledge, \methodname is the first bi-direction temporal network by focusing on multi-scale fusion in the SER, rather than simply concatenating forward and backward hidden states. 
\emph{Third}, we design a dynamic fusion module by combining dynamic receptive fields for learning the inter-dependencies at different temporal scales, so as to improve the model generalizability. Due to the articulation speed and pause time varying significantly across speakers, the speech requires different efficient receptive fields (\ie the time scale that reflects the affective characteristics) for each low-level feature (\eg MFCC).

\section{PROPOSED METHOD}
\label{METHODOLOGY}

\subsection{Input Pipeline}

To illustrate the temporal modeling capacity of our \methodname, we use the most commonly-used Mel-Frequency Cepstral Coefficients (MFCCs) features~\cite{IEMOCAP_CNN1} as the inputs to \methodname. We first set the sampling rate to the 22.050 kHz of each corpus and apply framing operation and Hamming window to each speech signal with 50-ms frame length and 12.5-ms shift. Then, the speech signal undergoes a mel-scale triangular filter bank analysis after performing a 2,048-point fast Fourier transform to each frame. Finally, each frame of the MFCCs is processed by the discrete cosine transformation, where the first 39 coefficients are extracted to obtain the low-frequency envelope and high-frequency details.

\begin{table*}[]
\renewcommand{\arraystretch}{1.0}
\centering
\small
\caption{The overall results of different SOTA methods on 6 SER corpora. Evaluation measures are UAR(\%) / WAR(\%). The `-' implies the lack of this measure. Furthermore, the superscripts indicate different evaluation settings. The `$^{*}$' implies a 10-fold cross-validation with 90\% and 10\% samples in train and test sets respectively, whose model is only evaluated at the last epoch. The `$^{**}$' implies a 10-fold cross-validation that 90\% of samples are used for training and 10\% for both validating and testing. Note that methods without superscript means that there is no source code to verify their specific experimental details.}

\begin{tabular*}{1\linewidth}{@{\extracolsep{\fill}}lcc|lcc|lcc}
\toprule[1.5pt]
\textbf{Model}    & \textbf{Year}     & \textbf{CASIA}         & \textbf{Model}   & \textbf{Year}      & \textbf{EMODB}         & \textbf{Model}   & \textbf{Year}      & \textbf{EMOVO}         \\
\midrule
DT-SVM~\cite{DT_SVM}  & 2019     & 85.08 / 85.08          & TSP+INCA~\cite{TSP_INCA} & 2021       & 89.47 / 90.09           & RM+CNN~\cite{RM_CNN} & 2021       & 68.93 / 68.93            \\
TLFMRF~\cite{TLFMRF} & 2020        & 85.83 / 85.83           & GM-TCN$^{**}$~\cite{GM_TCNet} & 2022       & 90.48 / 91.39            & SVM~\cite{MFMC_SVM} & 2021        & 73.30 / 73.30          \\
GM-TCN$^{**}$~\cite{GM_TCNet} & 2022       & 90.17 / 90.17          &  LightSER$^{**}$~\cite{IEMOCAP_CNN2}   & 2022    & 94.15 / 94.21        & TSP+INCA~\cite{TSP_INCA} & 2021       & 79.08 / 79.08           \\
CPAC$^{**}$~\cite{CPAC_IJCAI}  & 2022        & 92.75 / 92.75           & CPAC$^{**}$~\cite{CPAC_IJCAI}  & 2022         & 94.22 / 94.95          & CPAC$^{**}$~\cite{CPAC_IJCAI}  & 2022       & 85.40 / 85.40           \\\midrule
\textbf{\methodname}$^{*}$ & 2023 & \textbf{91.08 / 91.08}&
\textbf{\methodname}$^{*}$ & 2023 & \textbf{89.19 / 90.28}&
\textbf{\methodname}$^{*}$ & 2023 & \textbf{86.56 / 86.56}\\
\textbf{\methodname}$^{**}$ & 2023 & 94.67 / 94.67& 
\textbf{\methodname}$^{**}$ & 2023 & 95.17 / 95.70&
\textbf{\methodname}$^{**}$ & 2023 & 92.00 / 92.00

\\\midrule\midrule
\textbf{Model}   & \textbf{Year}      & \textbf{IEMOCAP}         & \textbf{Model}    & \textbf{Year}     & \textbf{RAVDESS}       & \textbf{Model}    & \textbf{Year}     & \textbf{SAVEE}     \\
\midrule
MHA+DRN~\cite{IEMOCAP_DRN}  & 2019     & 67.40 / ~~~~-~~~~            & CNN+INCA~\cite{INCA_CNN} & 2021        & ~~~~-~~~~ / 85.00          & DCNN~\cite{DCNN} & 2020       & ~~~~-~~~~ / 82.10           \\
CNN+Bi-GRU~\cite{IEMOCAP_GRU}  & 2020     & 71.72 / 70.39          & TSP+INCA~\cite{TSP_INCA} & 2021        & 87.43 / 87.43           &    TSP+INCA~\cite{TSP_INCA} & 2021        & 83.38 / 84.79         \\
SPU+MSCNN~\cite{IEMOCAP_CNN1} & 2021   & 68.40 / 66.60             & GM-TCN$^{**}$~\cite{GM_TCNet} & 2022       & 87.64 / 87.35           &     CPAC$^{**}$~\cite{CPAC_IJCAI}  & 2022         & 83.69 / 85.63           \\
LightSER$^{**}$~\cite{IEMOCAP_CNN2}   & 2022    & 70.76 / 70.23           & CPAC$^{**}$~\cite{CPAC_IJCAI}  & 2022         & 88.41 / 89.03          & GM-TCN$^{**}$~\cite{GM_TCNet} & 2022       & 83.88 / 86.02        \\\midrule
\textbf{\methodname}$^{*}$ & 2023 & \textbf{69.00 / 68.29}&
\textbf{\methodname}$^{*}$ & 2023 & \textbf{90.04 / 90.07}&
\textbf{\methodname}$^{*}$ & 2023 & \textbf{77.26 / 79.36}\\
\textbf{\methodname}$^{**}$ & 2023 & 72.50 / 71.65&
\textbf{\methodname}$^{**}$ & 2023 & 91.93 / 92.08&
\textbf{\methodname}$^{**}$ & 2023 & 86.07 / 87.71\\\bottomrule[1.5pt]

\end{tabular*}
\label{tab:SOTA}
\end{table*}

\subsection {Temporal-aware Bi-direction Multi-scale Network}

We propose a novel temporal emotional modeling approach called \methodname, which learns long-range emotional dependencies from the forward and backward directions and captures multi-scale features at frame-level. Fig.~\ref{fig:architecture} presents the detailed network architecture of \methodname. For learning multi-scale representations with long-range dependencies, the \methodname consists of $n$ Temporal-Aware Blocks (TABs) in both forward and backward directions with different temporal receptive fields. Next, we detail each component.

\noindent\textbf{Temporal-aware block.}\quad We design the TAB to capture dependencies between different frames and automatically select the affective frames, severing as a core unit of \methodname. As shown in Fig.~\ref{fig:architecture}, $\mathcal{T}$ denotes a TAB, each of which consists of two sub-blocks and a sigmoid function $\sigma(\cdot)$ to learn temporal attention maps $\mathcal{A}$, so as to produce the temporal-aware feature $\mat{F}$ by element-wise production of the input and $\mathcal{A}$. For the two identical sub-blocks of the $j$-th TAB $\mathcal{T}_j$, each sub-block starts by adding a DC Conv with the exponentially increasing dilated rate $2^{j-1}$ and causal constraint. 
The dilated convolution enlarges and refines the receptive field and the causal constraint ensures that the future information is not leaked to the past. 
The DC Conv is then followed by a batch normalization, a ReLU function, and a spatial dropout.

\noindent\textbf{Bi-direction temporal modeling.}\quad To integrate complementary information from the past and the future for the judgement of emotion polarity and modeling long-range temporal dependencies, we devise a novel bi-direction architecture based on the multi-scale features as shown in Fig.~\ref{fig:architecture}.
Formally, for the $\vec{\mathcal{T}}_{j+1}$ in the forward direction with the input $\vec{\mat{F}}_{j}$ from previous TAB, the output $\vec{\mat{F}}_{j+1}$ is given by Eq.~\eqref{eq:forward}:
\begin{align}
    \vec{\mat{F}}_{j+1}= \mathcal{A}(\vec{\mat{F}}_{j})\odot \vec{\mat{F}}_{j},\label{eq:forward}\\
    \cev{\mat{F}}_{j+1}= \mathcal{A}(\cev{\mat{F}}_{j})\odot \cev{\mat{F}}_{j},\label{eq:back}
\end{align}
where $\vec{\mat{F}}_{0}$ comes from the output of the first $1\times 1$ Conv layer and the backward direction can be defined similarly in Eq.~\eqref{eq:back}.

We then combine bidirectional semantic dependencies and compact global contextual representation at utterance level to perceive context as follows:
\begin{align}
    \vct{g}_j&=\mathcal{G}(\vec{\mat{F}}_{j}+\cev{\mat{F}}_{j}) \label{equ:GAP_feat},
\end{align}
where the global temporal pooling operation $\mathcal{G}$ takes an average over temporal dimension, yielding a representation vector for one specific receptive field from the $j$-th TAB.

\noindent\textbf{Multi-scale dynamic fusion.}\quad Furthermore, since the pronunciation habits (\eg speed or pause time) vary from speaker to speaker, the utterances have the characteristics of temporal scale variation. SER benefits from taking dynamic temporal receptive fields into consideration. We design the dynamic fusion module to adaptively process speech input at different scales, aiming to determine suitable temporal scale for the current input during the training phase. We adopt a weighted summation operation to fuse the features with Dynamic Receptive Fields (DRF) fusion weights $\mat{w}_\text{drf}$ from different TABs. The DRF fusion is defined as follows: 
\begin{align}
    \vct{g}_\text{drf}=\sum\nolimits_{j=1}^{n}w_j\vct{g}_j,  \label{equ:dynamic_rf}
\end{align}
where $\mat{w}_\text{drf} = [w_1,w_2,\ldots,w_{n}]\T$ are trainable parameters. 

Once the emotional representation $\mat{w}_\text{drf}$ is generated with great discriminability, we can simply use one fully-connected layer with the softmax function for emotion classification.

\section{EXPERIMENTS}
\label{EXPERIMENTS}

\subsection{Experimental Setup}
\noindent\textbf{Datasets.}\quad
To demonstrate the effectiveness of the proposed \methodname, we compare \methodname with State-Of-The-Art (SOTA) methods on 6 benchmark SER corpora. 
CASIA~\cite{CASIA} is a Chinese corpus collected from 4 Chinese speakers exhibiting 6 emotional states. EMODB~\cite{EMODB} is a German corpus that covers 7 emotions by 10 German speakers. EMOVO~\cite{EMOVO} is an Italian corpus recorded by 6 Italian speakers simulating 7 emotional states. IEMOCAP~\cite{IEMOCAP} is an English corpus that covers 4 emotions from 10 American speakers. RAVDESS~\cite{RAVDE} is an English corpus of 8 emotions by 24 British speakers. SAVEE~\cite{SAVEE} is an English corpus recorded by 4 British speakers in 7 emotions. 

\begin{figure*}[t]
	\centering
	\includegraphics[width=1.0\textwidth]{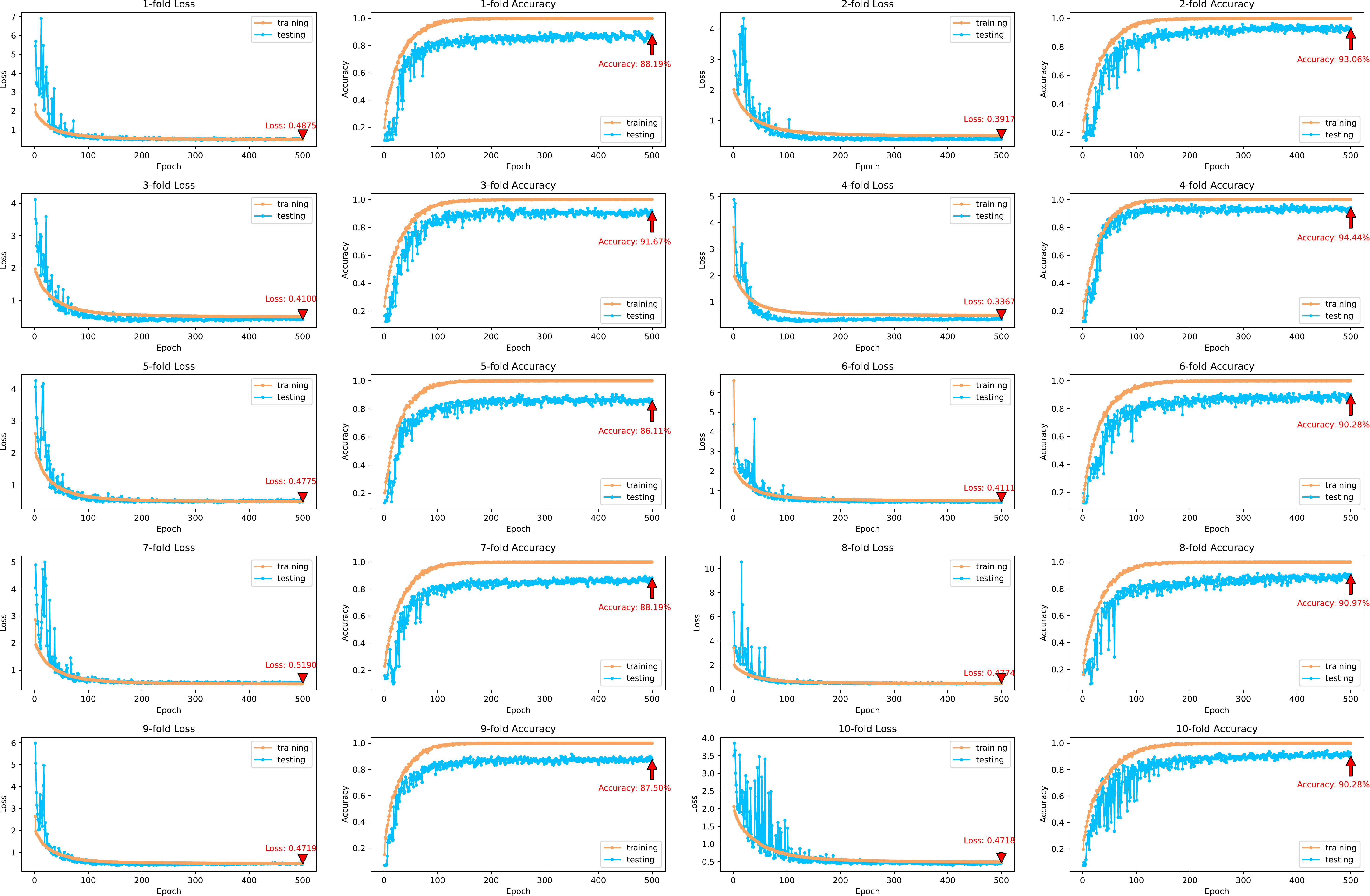}
	\caption{The accuracy and loss curves for 10-fold cross validation$^{*}$ on the RAVDESS corpus.}
	\label{fig:log}
\end{figure*}

\noindent\textbf{Implementation details.}\quad
In the experiments, 39-dimensional MFCCs are extracted from the Librosa toolbox~\cite{librosa}. The cross-entropy criterion is used as the objective function and the overall epoch is set to 500. Adam algorithm is adopted to optimize the model with an initial learning rate $\alpha$ = $0.001$, and a batch size of 64. To avoid over-fitting during the training phase, we implement label smoothing with factor 0.1 as a form of regularization. For the $j$-th TAB $\mathcal{T}_j$, there are 39 kernels of size 2 in Conv layers, the dropout rate is 0.1, and the dilated rate is $2^{j-1}$. To guarantee that the maximal receptive field covers the input sequences, we set the number of TAB $n$ in both directions to 10 for IEMOCAP and 8 for others. 
For fair comparisons with the SOTA approaches in experiments, following previous works~\cite{TSP_INCA,IEMOCAP_CNN2,IEMOCAP_DRN}, we mainly perform 10-fold cross-validation (CV) with 90\% training data and 10\% testing data in one fold to evaluate fitting ability of the model. To evaluate the generalization ability of the model, we further conduct experiments on six corpora under another evaluation setting. As shown in Table~\ref{tab:SOTA}, the superscript `$^{*}$' implies a 10-fold CV with 90\% and 10\% samples in train and test sets, whose model is only evaluated at the last epoch using the testing set.

\noindent\textbf{Evaluation metrics.}\quad Due to the class imbalance, we use two widely-used metrics, Weighted Average Recall (WAR) (\ie accuracy) and Unweighted Average Recall (UAR), to evaluate the performance of each method. WAR uses the class probabilities to balance the recall metric of different classes while UAR treats each class equally.

\subsection{Results and Analysis}
\noindent\textbf{Comparison with SOTA methods.}\quad
To demonstrate the effectiveness of our approach on each corpus, we select representative approaches on each corpus following the 10-fold CV strategy.  
Table~\ref{tab:SOTA} presents the overall results on 6 corpora, showing that our method significantly and consistently outperforms all these compared methods by a large margin. Remarkably, our approach gains 2.34\% and 2.61\% improvements of the average UAR and WAR scores than the second-best on each corpus under the second evaluating setting. However, most previous methods focus on evaluating the fitting ability of the model, leading to overfitting issues. 
We further evaluate the generalization ability of the model under another evaluation setting. As shown in Table~\ref{tab:SOTA}, although performance has declined, the \methodname still has competitive performance and good generalization ability on several corpora. Fig.~\ref{fig:log} shows that \methodname does not exhibit significant overfitting issues, and its convergence curves remain relatively stable. 
Moreover, it can be observed that the affective discrimination ability of \methodname in short-term speech (\eg CASIA, EMODB, EMOVO, and RAVDESS) is generally stronger than that in long-term speech (\eg IEMOCAP and SAVEE), which means that long-term dependence is still a challenging issue. 
Please refer to our GitHub repo\footnote{\url{https://github.com/Jiaxin-Ye/TIM-Net_SER}} for extra experimental details and results.

\begin{figure}[t]
	\centering
	\includegraphics[width=1.0\linewidth]{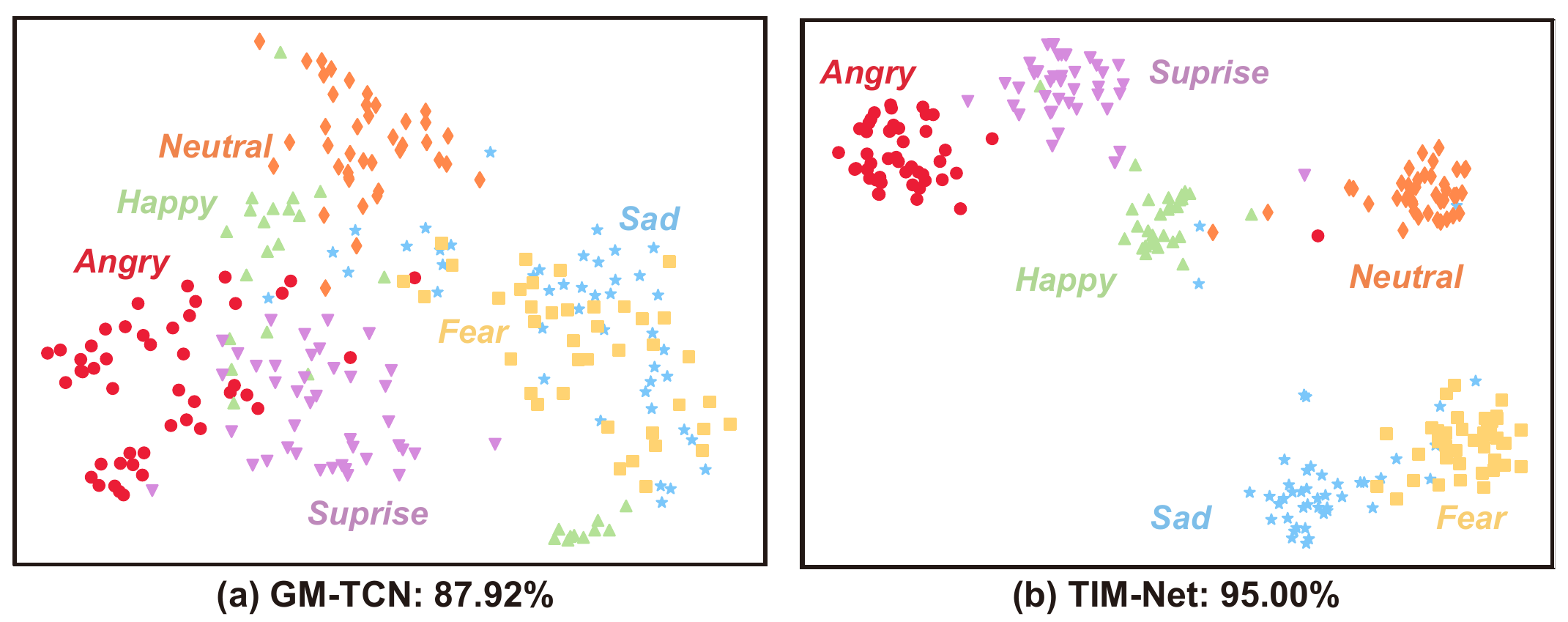}
	\caption{t-SNE visualizations of features learned from SOTA method GM-TCN and \methodname. The score denotes WAR.}
	\label{fig:tsne}
\end{figure}

\begin{table}[]
\renewcommand{\arraystretch}{1.0}
\small
\centering
\caption{UAR and WAR on the cross-corpus SER task with different methods. All values are the average $\pm$ std of 10 runs, each of which consists of 20 cross-corpus cases.}
\begin{tabular*}{1\linewidth}{@{\extracolsep{\fill}}cccc}
\toprule[1.5pt]
\addline{\textbf{Method}}  & \textbf{TCN}         & \textbf{CAAM}~\cite{CPAC_IJCAI}                        & \textbf{\methodname}\\\midrule
\addline{\textbf{\tabincell{\boldmath $\mathrm{UAR}_\mathrm{avg} \pm \mathrm{std}$\\\boldmath $\mathrm{WAR}_\mathrm{avg} \pm \mathrm{std}$}}} & \tabincell{24.47 $\pm$ 0.38\\24.39 $\pm$ 0.42}                     & \tabincell{32.37 $\pm$ 0.27\\33.65 $\pm$ 0.41}                         & \textbf{\tabincell{34.49 $\pm$ 0.43\\35.66 $\pm$ 0.32}}\\
\bottomrule[1.5pt]
\end{tabular*}
\label{tab:generalization}
\end{table}

\noindent\textbf{Visualization of learned affective representation.}\quad
To investigate the impact of \methodname on representation learning, we visualize the representations learned by \methodname and GM-TCN~\cite{GM_TCNet} through the t-SNE technique~\cite{T-SNE} in Fig.~\ref{fig:tsne}. 
For a fair comparison, we first use the same 8:2 hold-out validation on CASIA corpus for the two methods, and visualize the representations of the same test data after an identical training phase. 
Although GM-TCN also focuses on multi-scale and temporal modeling. Fig.~\ref{fig:tsne}(a) shows heavy overlapping between \emph{Fear} and \emph{Sad} or \emph{Angry} and \emph{Surprise}.
In contrast, Fig.~\ref{fig:tsne}(b) shows that the different representations are clustered with clear classification boundaries. 
The results confirm that the \methodname provides more class-discriminative representations to support superior performance by capturing intra- and inter-dependencies at different temporal scales.

\noindent\textbf{Domain generalization analysis.}\quad
Due to various languages and speakers, the SER corpora, although sharing the same emotion, have considerably significant domain shifts. The generalization of the model to unseen domain/corpus is critically important for SER. Inspired by the domain-adaptation study in CAAM~\cite{CPAC_IJCAI}, we likewise validate the generalizability of \methodname on the cross-corpus SER task, following the same experimental setting as CAAM except that \methodname does not have access to the target domain. Specifically, we likewise choose 5 emotional classes for a fair comparison, \ie \emph{angry}, \emph{fear}, \emph{happy}, \emph{neutral}, and \emph{sad}, shared among these 5 corpora (except for IEMOCAP, which has only 4 emotions). These 5 corpora form 20 cross-corpus combinations. And we report the average UAR and WAR, and their standard deviation from 10 random runs for each task in Table~\ref{tab:generalization}. 

The performance of TCN over different corpora is close to random guessing with odds equal to 25\%, and \methodname has a significant improvement over TCN. Surprisingly, \methodname outperforms CAAM, one latest task-specific domain-adaptation method. The results suggest that our \methodname is effective in modeling emotion with strong generalizability.

\subsection{Ablation Study}
We conduct ablation studies on all the corpus datasets, including the following variations of \methodname: \textbf{TCN}: the \methodname is replaced with TCN; \textbf{w/o BD}: the backward TABs are removed while keeping the forward TABs; \textbf{w/o MS}: the multi-scale fusion is removed and $\vct{g}_n$ is used as $\vct{g}_{\text{drf}}$ corresponding to max-scale receptive field; \textbf{w/o DF}: the average fusion is used to confirm the advantages of dynamic fusion. The results of ablation studies are  shown in Table~\ref{tab:ablation}. We have the following observations. 

\emph{First}, all components contribute positively to the overall performance. 
\emph{Second}, our method achieves 8.31\% and 8.41\% performance gains in UAR and WAR over TCN that also utilizes DC Conv. Since the inability of TCN to capture contextual multi-scale features, capturing intra- and inter-dependencies at different temporal scales is critical to SER.
\emph{Third}, when removing the backward TABs or multi-scale strategy, the results substantially drop due to the weaker capacity to model temporal dependencies and perceive the sentimental features with different scales. 
\emph{Finally}, \methodname without dynamic fusion performs worse than \methodname, which verifies the benefits of deploying dynamic fusion to adjust the model adaptively. 
\begin{table}[]
\renewcommand{\arraystretch}{1.0}
\small
\centering
\caption{The average performance of ablation studies and \methodname under 10-fold CV on all six corpora. The `w/o' means removing the component from \methodname}
\begin{tabular*}{1\linewidth}{@{\extracolsep{\fill}}lccccc}
\toprule[1.5pt]
\addline{\textbf{Method}}  & \textbf{\tabincell{TCN} } & \textbf{\tabincell{w/o BD}} & \textbf{\tabincell{w/o MS}} & \textbf{\tabincell{w/o DF}} & \textbf{\tabincell{\methodname}}\\\midrule
\addline{\tabincell{\boldmath $\mathrm{UAR}_\mathrm{avg}$\\\boldmath $\mathrm{WAR}_\mathrm{avg}$}} & \tabincell{80.45\\80.56}             & \tabincell{84.92\\85.32}               & \tabincell{85.45\\85.82}              & \tabincell{84.85\\85.24}                   & \tabincell{\textbf{88.76}\\\textbf{88.97}}\\
\bottomrule[1.5pt]

\end{tabular*}
\label{tab:ablation}
\end{table}


\section{CONCLUSIONS}
\label{CONCLUSIONS}
In this paper, we propose a novel temporal emotional modeling approach, termed~\methodname, to learn multi-scale contextual affective representations from various time scales. 
\methodname can capture long-range temporal dependency through bi-direction temporal modeling and fuse multi-scale information dynamically for better adaptation to temporal scale variation. 
Our experimental results indicate that learning representation from the context information with dynamic temporal scales is crucial for the SER task. The ablation studies, visualizations, and domain generalization analysis further confirm the advantages of \methodname. 
In the future, we will investigate the disentanglement of emotion and speech content through the proposed temporal modeling approach for better generalization in cross-corpus SER tasks.

\vfill\pagebreak

{\small
\balance

}
\end{document}